**$CO_2$ laser-driven reactions in pure acetylene flow**

Peter V. Pikhitsa*, Daegyu Kim, and Mansoo Choi

*National CRI Center for Nano Particle Control, Institute of Advanced, Machinery and Design, School of Mechanical and Aerospace Engineering, Seoul National University, 151-742, Seoul, Korea*

*Corresponding author. Tel/Fax: +82 8769396 / + 82 8878762.

*E-mail address*: peter@snu.ac.kr

**Abstract**

We show that multiple-photon absorption of radiation from a 10.56 μm cw $CO_2$ laser by intermediates (ethylene, vinylidene) generated in pure acetylene flow makes them decompose to carbon dimers and excited hydrogen. The latter associates with downstream acetylene to feedback those laser absorbing intermediates thus making the reactions self-sustained in the absence of oxygen. This process is different from acetylene self-decomposition that may occur at higher temperature and pressure. The results of our work may be useful for understanding the generation of various carbon allotropes and interstellar dust from acetylene.

## I. INTRODUCTION

Currently there is a growing interest in generation of allotropic carbon solids, polycyclic aromatic hydrocarbons (PAH) and aliphatic hydrocarbons which are found in interstellar space [1-3] along with fullerenes discovered in a planetary nebula [4]. Recently it was found [5] that a light absorption feature in acetylene plasma may coincide with one broad diffuse interstellar band which induces one to take a close look to possible reactions in pure acetylene gas.

It was reported earlier [6] that pure acetylene flow inside a diffusion $O_2/H_2$ flame generated carbon shell-shaped nanoparticles when irradiated with a 10.56 μm cw $CO_2$ laser in spite of the fact that acetylene practically does not absorb the laser frequency [7]. The generation phenomenon was ascribed in Ref. [6] to the pre-heating of acetylene by $O_2/H_2$ flame, the initial pyrolysis and the incipient nanoparticles due to the heating, and then to the surface reactions sped up by additional heating of the incipient nanoparticles by the laser radiation.

However, our later experiments [8] demonstrated that the process of nanoparticle generation successfully occurs (after initialisation) in the absence of $O_2/H_2$ flame in pure acetylene flow shielded from air and diluted by inert gases (Ar, $N_2$, He) and $H_2$; moreover, many carbon allotropes were found among the nanoparticles, including nanodiamonds [8] (Fig. 1) that could not stand the high temperature in $O_2/H_2$ flame and thus were not observed in Ref. [6]. The laser power needed for the generation of shell-shaped nanoparticles turned out to be several times less than found in Ref. [6]. Additionally, PAH are produced and detected in ultraviolet (UV) absorption spectra from *extra situ* material [3].

The generation of nanoparticles is followed by blazing light from the reaction zone with a clearly non-Planckian broad spectrum in near UV region that tells us about non-thermal way of the reaction. It is useful to mention the review [9] describing allothermal effects that usually lead to diverse carbon structures. In Ref. [8] it was preliminary concluded that the intermediates (e.g. excited ethylene and vinylidene) that absorb the laser radiation [10-12] can keep the reaction going in laser-driven chain reactions aside from the soot-generating reactions; the latter can make acetylene be capable of sustaining a thermal self-decomposition flame only at high pressures and temperatures above $1200^0$ C.

Here we report on a theoretical description of the laser-driven reactions and compare the results with our experiment where a 10.56 μm cw $CO_2$ laser beam irradiates pure (or diluted with inert gases) acetylene flow shielded from air completely by nitrogen or argon (Fig. 2). The experimental set-up was the same as described in Ref. [6] with the exception that no $O_2/H_2$ flame was used (see Fig. 2a for the experimental setup). We additionally experiment with the gas flow rate and the dilution of acetylene.

The relevant parameters are the laser power $W$, the acetylene flow velocity $V$ and the acetylene concentration $n=xn_0$, where $x$ is the dilution ratio ($x$=1 for pure acetylene) and $n_0$ is the concentration of pure acetylene. We calculate the threshold values of the parameters at which the reaction terminates. Possible thermal acetylene decomposition and surface reactions are ignored in our approach as far as our thermocouple measurements of the gas flow temperature as close as only 2 mm above/below the reaction zone indicated the temperature not exceeding $500^0$ C. This fact excludes the thermal decomposition effects in acetylene at least for not too high laser power.

## II. THEORETICAL

The most general and most simplified reaction scheme for generation laser-absorbing intermediates may be the following:

$$C_2H_2 + H_2^* \rightarrow C_2H_4^*. \quad (1)$$

This general addition reaction (and its variations with participation of charged reagents, of atomic hydrogen *etc* [13-16]) occurs with the addition reaction rate constant $\alpha$. The asterisk denotes excited species. Then $C_2H_4^*$ decomposes under the $CO_2$ laser irradiation into $CH_2$=C: +$H_2$ and further, vinylidene decomposes into $C_2$ and $H_2^*$ as was described in Refs. [9-12]. Thus the parent $H_2^*$ molecule produces two molecules in the branching:

$$C_2H_4^* + hv \rightarrow C_2 + 2H_2^*. \quad (2)$$

The system of differential equations that follows from (1) and (2) predicts the critical behaviour with the critical gas velocity $V_c$ for given acetylene concentration that can terminate the reaction even at arbitrary high laser power. The exact solution of the corresponding equations gives the relative position $z$-$z_0$ of the reaction zone to be compared with the experimental one (Fig. 2b). From the solution we obtained for $z$ an approximation formula suitable for high laser power:

$$z=z_0 + (4D_h / V_c)(1-V/V_c)^{-1/2}, \quad (3)$$

where $D_h$ is the hydrogen diffusion coefficient;

$$V_c = 2\sqrt{\alpha n D_h} = 2\sqrt{\alpha n_0 D_h x} = V_1\sqrt{x} \ . \quad (4)$$

Let us derive Eqs. (3), (4). The reactions in Eqs. (1) (2), can be described by the following one dimensional steady-state differential equations (prime denotes differentiation on $z$):

$$V n_a' = -\alpha n_a n_h - \gamma n_a + D_a n_a'', \quad (5)$$

$$V n_e' = \alpha n_a n_h - \gamma n_e - w(z) n_e + D_e n_e'' \ , \quad (6)$$

$$V n_h' = -\alpha n_a n_h - \gamma n_h + 2w(z) n_e + D_h n_h'' \ , \quad (7)$$

where

$$w(z) = \begin{cases} w & z \geq 0 \\ 0 & z < 0 \end{cases} . \quad (8)$$

Here $w$ (an increasing function of the laser power $W$ [10,12]) is the rate of laser induced decomposition of the infrared-absorbing intermediate (with concentration $n_e$ and diffusion coefficient $D_e$) which we will conventionally call "ethylene" though, as we said above, vinyl radical and vinylidene can participate as well; $n_a$, $D_a$ and $n_h$, $D_h$ - concentrations and diffusion coefficients of acetylene and hydrogen, respectively; $V$ is the gas velocity in the direction of the vertical coordinate $z$; prime denotes the derivative with respect to $z$; $\gamma$ describes losses in the real system. Eq (8) describes the half-infinite laser beam which simplification allows analytical results. Pressure effects are not taken into account here, as far as the $C_2$ vapor quickly condenses into carbon nanoparticles in the reaction zone and thus the number of gas molecules can be considered nearly constant.

Boundary conditions are:

a) at $z = -\infty$, $n_a = n$, $n_e = 0$, $n_e = 0$;

b) at $z = \infty$, $n_a = 0$, $n_e = 0$, $n_h = n$.

c) the concentrations are continuous at $z = 0$.

If we neglect losses ($\gamma$=0), neglect diffusion of acetylene and ethylene (by setting $D_a = 0$ and $D_e = 0$ as far as hydrogen diffusion coefficient $D_h \approx 1\,\mathrm{cm}^2$/s is at least 10 times larger) and make substitutions $n_a = n\exp(-\varphi)$ and $n_h = (V/\alpha n)n\varphi'$ then Eqs. (5)-(7) can be combined in one equation:

$$D\varphi''' + (\omega D - v)\varphi'' - \varphi'\exp(-\varphi) - \omega v\varphi' + \omega[1-\exp(-\varphi)] = 0, \quad (9)$$

where $D = D_h/(\alpha n)$, $v = V/(\alpha n)$, $\omega = w/V$; then $n_e = n\left(D\varphi'' - v\varphi' + 1 - e^{-\varphi}\right)/2$. Note that these parameters have dimensions: [$D$]=cm$^2$, [$v$]=cm, and [$\omega$]=1/cm.

By direct substitution one can make sure that when $\omega v = 1/2$ and $v = \sqrt{D/2}$ (therefore $\omega D - v = 0$) Eq. (9) has the exact solution:

$\varphi = 2\ln\{1+\exp[(z-z_0)/\sqrt{2D}]\}$, (10)

Then for the concentrations one obtains:

$$n_a = \frac{n}{4}\left[1-\tanh\left(\frac{z-z_0}{2\sqrt{2D}}\right)\right]^2 ;\ n_h = \frac{n}{2}\left[1+\tanh\left(\frac{z-z_0}{2\sqrt{2D}}\right)\right]; n_e = \frac{n}{4}\cosh^{-2}\left(\frac{z-z_0}{2\sqrt{2D}}\right). \ (11)$$

The presence of $z_0$ means that there is an arbitrary position of the soliton solutions in Eq. (11). Thus the reaction zone can be "blown up" arbitrarily high and parameters $\omega v = 1/2$ and $v = \sqrt{D/2}$ are therefore critical for the reaction. Returning to dimensional parameters one gets $V_{thr} = \sqrt{\alpha n D_h / 2}$ and $w_{thr} = w_0 = \alpha n / 2$.

Let us show that the parameter values obtained above are on the critical curve which separates reaction/no reaction regions in the parameter domain. The curve can be found from Eqs. (5)-(7) after their linearization at $(n-n_a), n_e, n_h << n$ with the solution in form of $\exp(\beta z - \beta z_0)$, $\beta > 0$, that corresponds to the asymptotic solution of the soliton type when the soliton is far on the right side from z=0, therefore for Eq. (11) $\frac{z_0 - z}{2\sqrt{2D}}$ >>1 and $n_e \approx n\exp[(z-z_0)/\sqrt{2D}]$ from Eq. (11). Thus, $\beta_{thr} = 1/\sqrt{2D} = \sqrt{\alpha n/(2D_h)}$ is found from the asymptotic of the exact solution of Eq. (11).

One finds from linearized Eq. (6), (7)

$$w = \frac{(-\beta V - \gamma + D_e\beta^2)(-\beta V - \gamma - \alpha n + D_h\beta^2)}{-\beta V - \gamma + \alpha n + D_h\beta^2} = \frac{(-\beta V)(-\beta V - \alpha n + D_h\beta^2)}{-\beta V + \alpha n + D_h\beta^2} \ (12)$$

as far as we neglected $\gamma$ and $D_e$ for simplification. One can make sure that for $\beta > 0$ not any positive values of $w$ are possible. At a given $V$ the value of $w$ which is a marginal value separating existing and non-existing solutions is given by the equation:

$$\frac{\partial w}{\partial \beta} = 0. \ (13)$$

Eq. (13) can be rewritten as:

$$V = \alpha n / \beta + D_h \beta - \sqrt{2\alpha n}\sqrt{\alpha n / \beta^2 - D_h} \quad . \text{ (13a)}$$

Both of Eqs. (12), (13a) define the critical diagram shown in Fig. 3a. The direct substitution of the parameter values ($\beta_{thr} = \sqrt{\alpha n/(2D_h)}$, $V_{thr} = \sqrt{\alpha n D_h / 2}$, and $w_{thr} = \alpha n / 2$) found above for the exact solution of Eq. (11) into Eqs. (12), (13a) satisfies them. Therefore, the soliton solution is indeed a threshold solution. As one can derive from Eqs. (12), (13), the critical diagram in Fig. 3a has the critical velocity of Eq. (4) $V_c = 2\sqrt{\alpha n D_h} = 2\sqrt{\alpha n_0 D_h x}$ (then $\beta_c = \sqrt{\alpha n / D_h}$ and $w_c = \infty$).

If the gas velocity overcomes the critical velocity, no laser power can support the reaction. This phenomenon is observed in our experiment (see Fig. 2). There is no reaction in the centre of the reaction zone when the velocity there is above $V_c$. The laser power was 900 W and the beam diameter was around 2 mm thus the power density of 7162 W/cm$^2$ can be considered as a high power density [12]. Note the blue-violet light [19] from the lower part of the reaction zone which may indicate acetylene/hydrogen plasma emission. Also there is the low power threshold for the laser power [12] as far as multi-photon processes are involved.

For the precise determination of the critical velocity one has to calculate the position of the reaction zone and to follow its change with the gas velocity or dilution. From computing Eqs.(5)-(8) for two different rates $w/w_0=1$ (exact solution) and $w/w_0=100$ we found the position of the reaction zone at velocities $V$ approaching the threshold velocities $V_{thr}$ (Fig. 3b,c). Using the boundary conditions at $z=0$ one can show that for the location of the reaction zone maximum $z$ there is an asymptotic $z=z_0 + A\, \beta_c^{-1} (1-V/V_{thr})^{-1/2}$. Dimensionless $A$, which is difficult to calculate analytically in a general case, may slightly vary with $w$. One can make a good approximation in Fig. 3d for two superposed dependencies calculated at very different $w$ with $A \approx 2$

$$z=z_0 + (4D_h / V_c)(1-V/V_{thr})^{-1/2} \text{ (14)}$$

which leads to Eq. (3) for high power as far as $V_{thr}=V_c$. The shift $z_0$ was adjusted to superpose data on one curve.

## III. COMPARISON WITH EXPERIMENT AND DISCUSSION

Let us compare Eq. (3) with the experimental data shown in Fig. 4a,b. The transformation coefficient from flow rate to gas velocity is 0.25 cm $s^{-1}$ $(ccm/min)^{-1}$ that is the critical flow of 338 ccm/min in Fig. 4b corresponds to $V_c$ = 84.5 cm/s at this dilution of $x$=0.6. Fitting the experimental dependency of the position $z$ of the reaction zone against the gas velocity at dilution $x$=0.6 ($x$=0.4) gives values of $D_h$ =2.5 $cm^2$/s ($D_h$ =2.87 $cm^2$/s) and $an_0$ = 1000 $s^{-1}$ ($an_0$ = 1150 $s^{-1}$) and therefore from Eq. (4) $V_1$ = 101cm/s ($V_1$ = 115cm/s). Additionally, Eq. (4) was used for fitting the velocities when the reaction zone starts breaking into halves (eye observation) at various dilutions which gives $V_1$ = 96.2 cm/s (see Fig. 4c). The calculated value $\beta_c$=2 $mm^{-1}$ gives additional opportunity to check our theory by measuring the light intensity spatial distribution near the edge of the reaction zone by using a digital camera.

We took images of the reaction zone at different shutter speeds thus controlling the position of the apparent reaction edge seen by the digital camera (Fig. 5). The plot of the shutter speed against the edge position determined from the images gives the distribution of the intensity of the reaction (Fig. 5) with the characteristic spatial decay parameter $\beta$=2.5 $mm^{-1}$. A reasonable agreement between theoretical values and the experimental ones supports the mechanism for the acetylene conversion to carbon through laser absorbing gaseous intermediates. Their decomposition and the following condensation of resulting carbon dimers at low-temperature conditions leads to the metastable carbon [8] including nanodiamonds. The metastable carbon rapidly transforms into continuous layer shell-shaped carbon nanoparticles or nanohorns [20] (Fig. 1) under the laser beam similar to [17] where crystallization in 50 milliseconds of a carbon nanotube from an amorphous nanotube was reported.

Note that the case of pure ethylene is different. First, no splitting of the reaction zone is observed as far as ethylene fuel is absorbing the laser power [8] and also the feedback scheme described by Eq. (1) does not work for pure ethylene. Second, it is seen that the reaction is supported by too extensive heating of the reaction zone with twice more hydrogen and the products which are mostly soot-like carbon [18] and PAH. Yet, acetylene might also be created in this process [11, 19] and then additionally participate in the reaction according to the mechanism described above. However the reaction intensity should be lower than for pure acetylene as we indeed observe in our experiment.

## IV. CONCLUSIONS

In conclusions, grounding on the theoretical description of our experiment we suggested steady state laser-driven reactions in pure acetylene flow that may be interesting for further investigation of number of processes in hydrocarbons: from tailored synthesis of carbon allotropes to photochemistry of hydrocarbons and the understanding of interstellar grains.

## Acknowledgements

Financial support from the Acceleration Research Program supported by the Korean Ministry of Science and Technology is gratefully acknowledged. Support from the BK21 program and WCU (World Class University) multiscale mechanical design program through the Korea Research Foundation is also gratefully acknowledged. P.V.P. thanks Fyodor Sirotkin for taking photos of the reaction zone.

Figure captions

FIG. 1. Transmission electron microscopy images of (a) shell-shaped carbon nanoparticle [6,8], (b) single-walled nanohorns (see [20]) collected from the carbon stream, (c) graphene sheets, and (d) nanodiamonds [8]. The inset shows the selected area diffraction pattern for nanodiamonds.

FIG. 2. (Color online) The interaction of the acetylene flow with the laser beam. (a) Experimental set-up, (b) The evolution of the reaction zone at different flow rates. The diameter of the inner nozzle is 3 mm. $Z$ is the position of the reaction depending on the gas velocity in the centre ($Z$=2.5 mm in the photo). Bottom and right panels show successive evolution of the reaction zone with the increasing flow rate from right to left and from top to bottom, respectively. The right panel corresponds to $x=0.4$ (see Fig. 4a).

FIG. 3. (Color online) The laser-driven reaction in acetylene: (a) The critical diagram for the reaction from Eqs. (12), (13). The area under the critical curve corresponds to no reaction. Arrows indicate the approach to the threshold velocities at two different values of the power; (b), (c) computer solutions of Eq. (5)-(8) for the concentration of intermediates at approaching the threshold velocity at the power of the exact soliton solution and at the power 100 times more, correspondingly; (d) the approximation of the "universal" position dependency.

FIG. 4. Experimental data on the reaction zone evolution with the change in the gas velocity. (a) The position of the reaction zone against the acetylene velocity at dilution ratio $x=0.4$ and (b) $x=0.6$. Squares are experimental; the line is the fitting by Eq. (3). (c) The critical flow for the interruption of the reaction zone vs. dilution (squares). The line is the fit with Eq. (4).

FIG. 5. (Color online) The reaction intensity profile measured indirectly by the position of the luminous reaction edge (lower part of the reaction zone) seen by the digital camera at different shutter speeds. Five data points (filled circles) correspond to five images on the left. The red line is the exponential fit which gives $\beta$=2.5 mm$^{-1}$ to be compared with $\beta_c$=2 mm$^{-1}$ that is calculated with $D_h$ =2.5 cm$^2$/s and $an_0$= 1000 s$^{-1}$ .

Figure 1

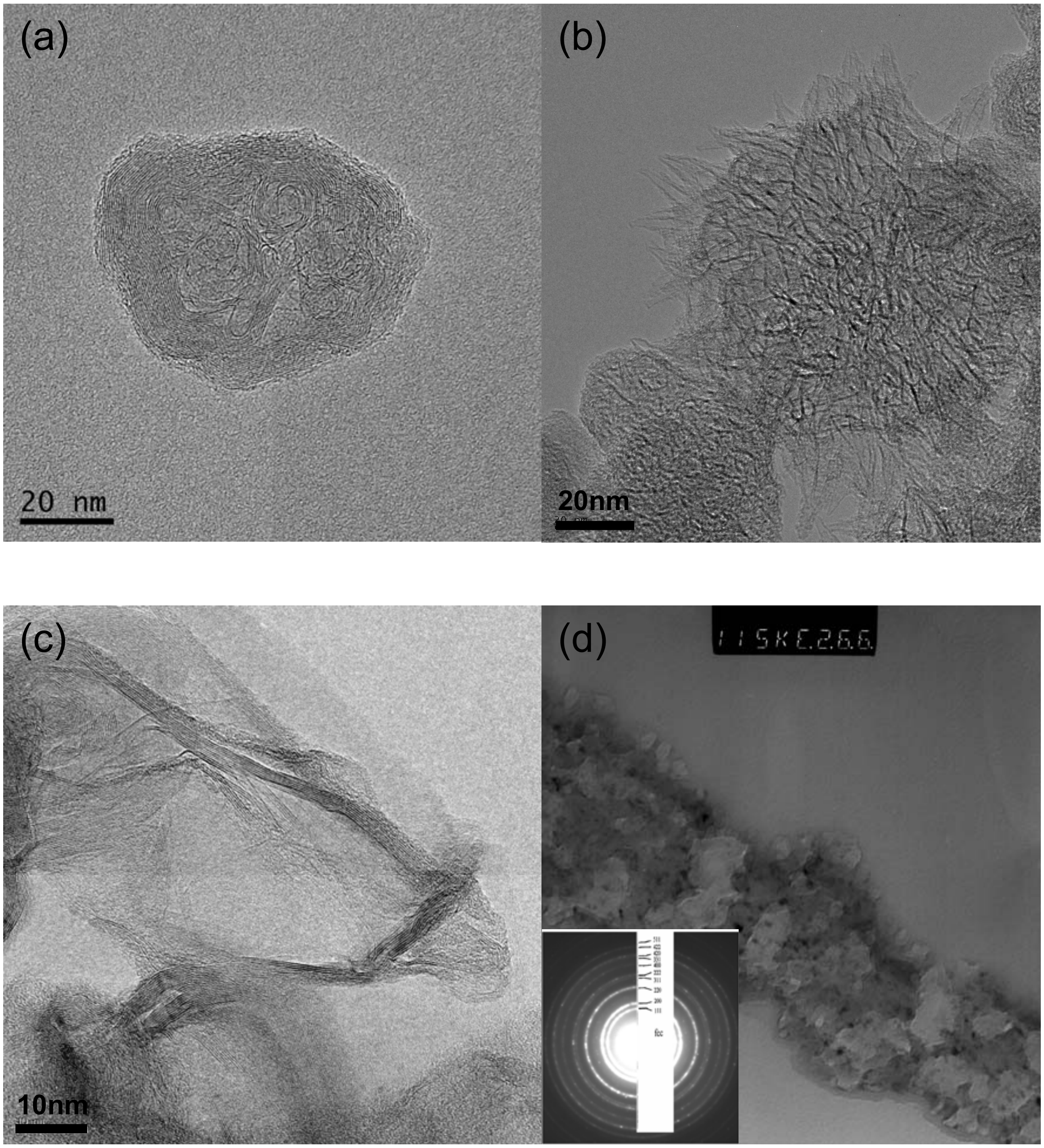

Figure 2

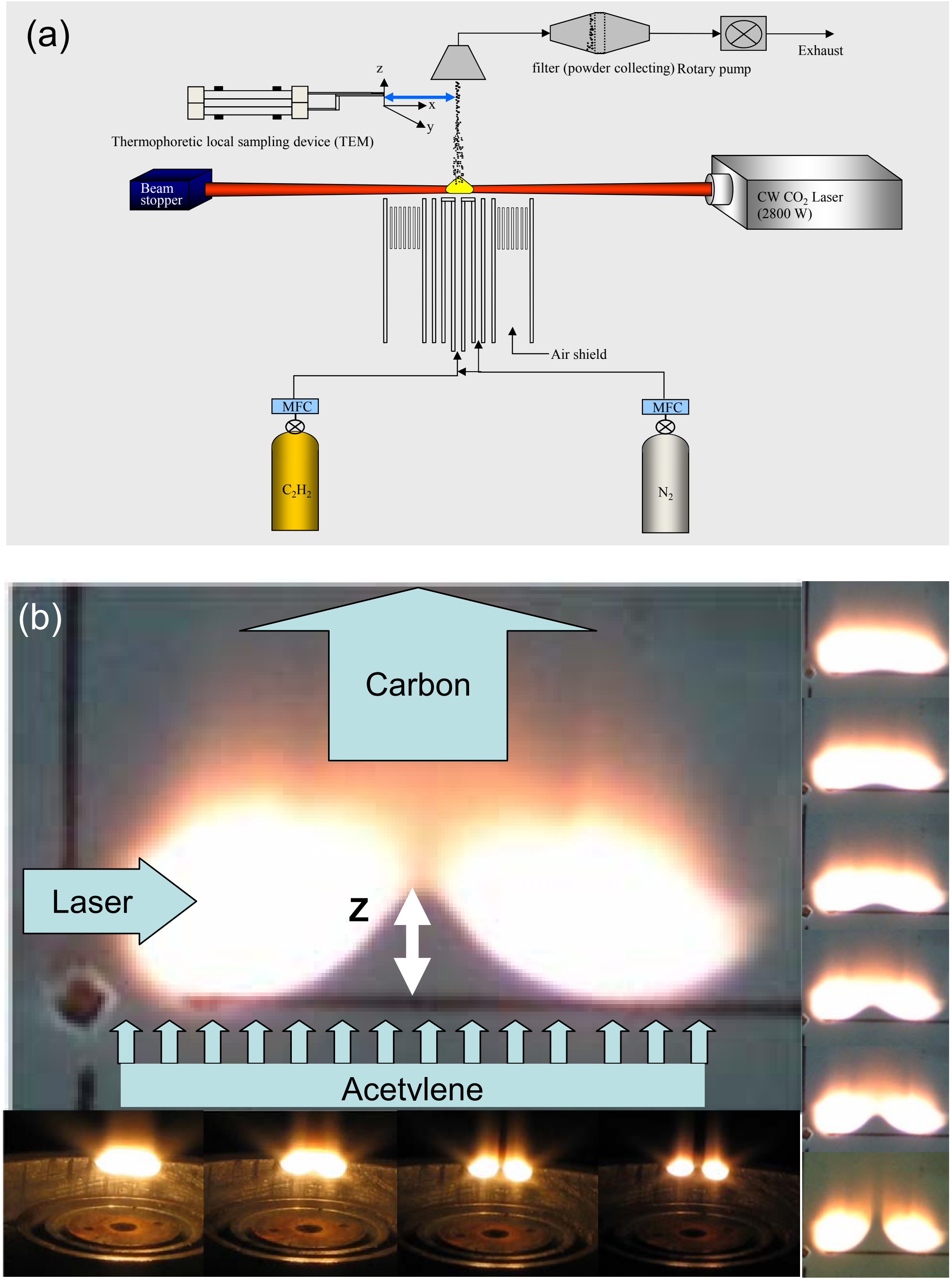

Figure 3

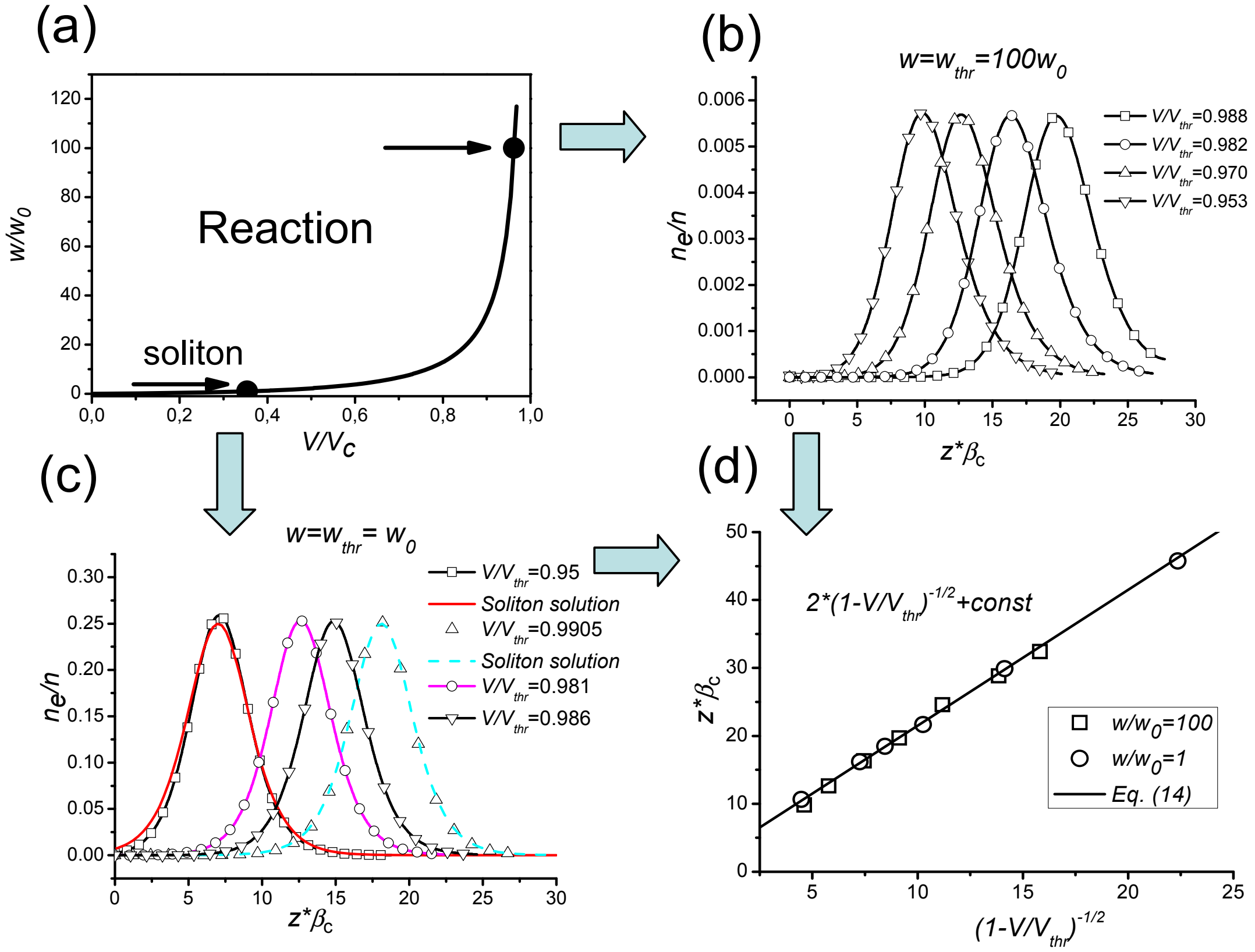

Figure 4

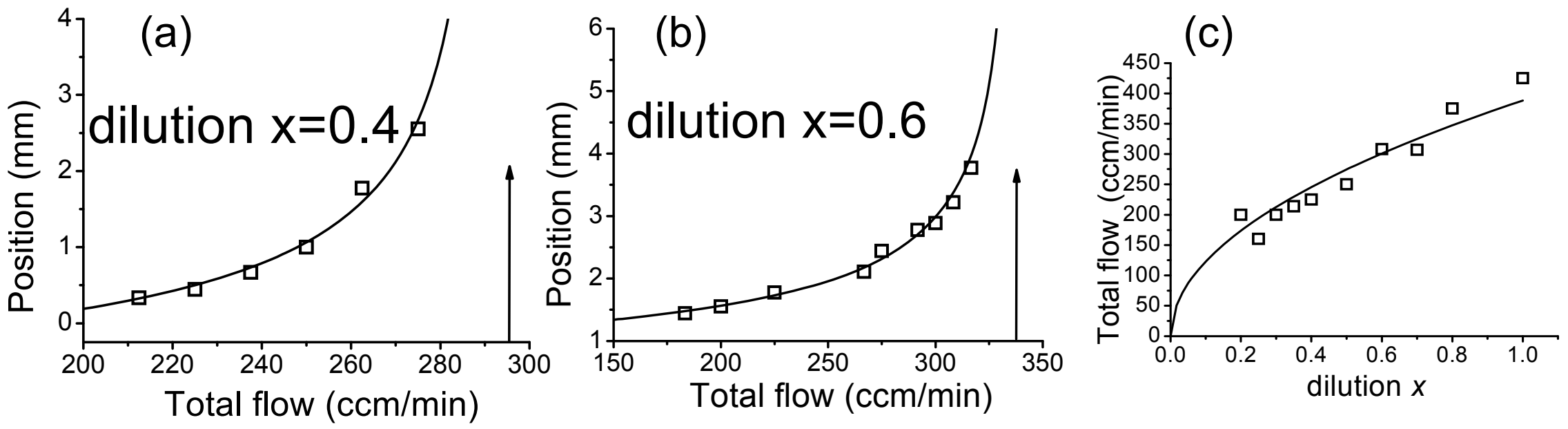

Figure 5

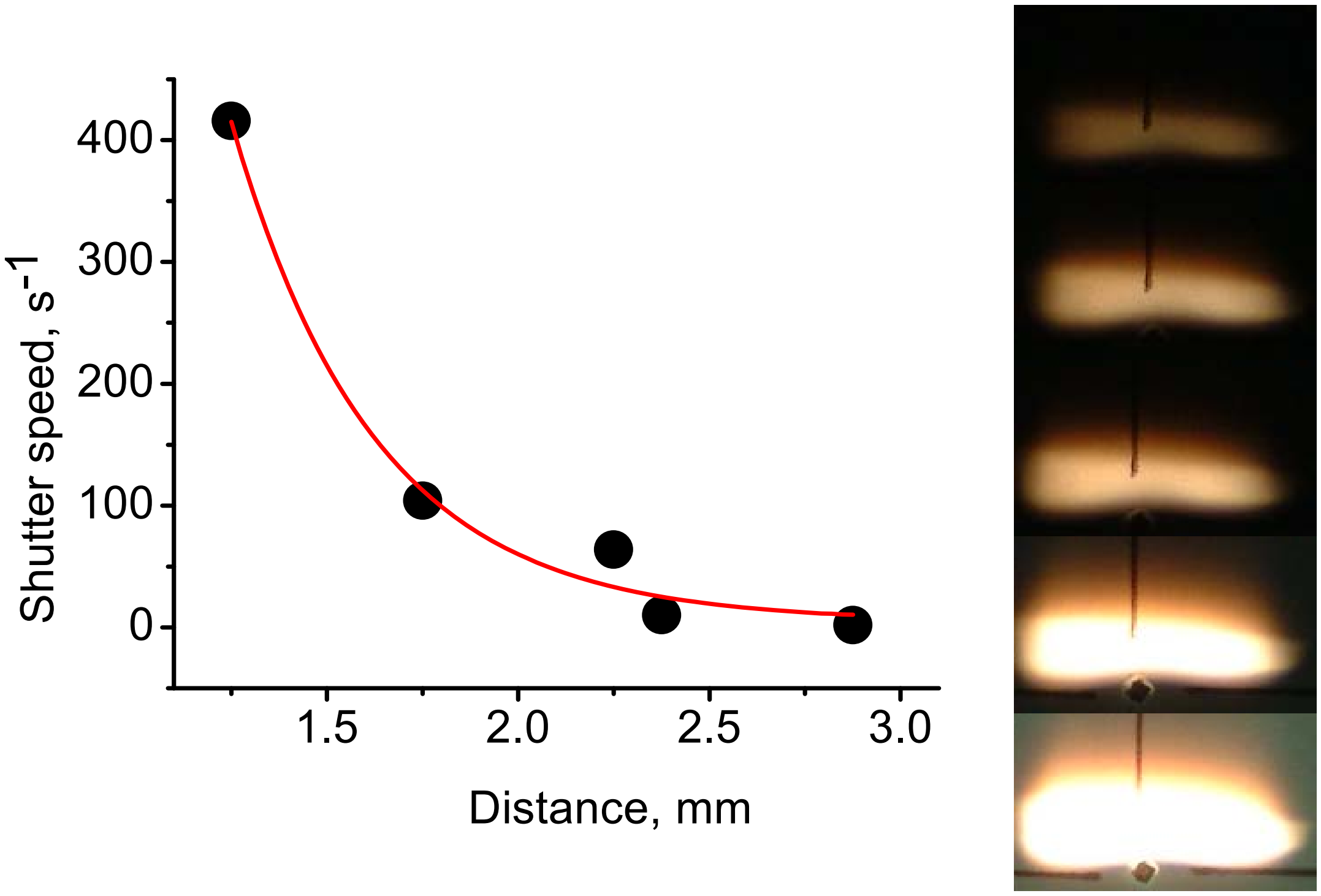